\documentclass[a4paper,twocolumn]{Gaia2004} 
\usepackage{times}      
\usepackage{epsfig}     
\usepackage{natbib}     
\title{The Payload Data Handling and Telemetry Systems of Gaia}

\author[1,2]{J. Portell}
\author[3,1]{X. Luri}
\author[1,2]{E. Garc\'ia-Berro}
\author[2]{E.M. Geijo}
\affil[1]{Institut d'Estudis Espacials de Catalunya, 
          c/Gran Capit\`a 2-4, 08034 Barcelona, Spain}
\affil[2]{Departament de F\'isica Aplicada, 
          Universitat Polit\`ecnica de Catalunya, 
          Avda. Canal Ol\'impic s/n, 
          08860 Castelldefels, Spain}
\affil[3]{Departament d'Astronomia i Meteorologia, 
          Universitat  de Barcelona, 
          c/ Mart\'{\i} i Franqu\`es 1, 
          08028 Barcelona, Spain}

\bibpunct{(}{)}{;}{a}{}{,}  

\begin{document}


\maketitle

\begin{abstract}
The Payload  Data Handling  System (PDHS) of  Gaia is  a technological
challenge, since it will have  to process a  huge amount of  data with
limited resources. Its main  tasks include the optimal codification of
science  data, its  packetisation  and its  compression, before  being
stored on-board  ready to  be transmitted. Here  we describe a  set of
proposals  for its  design, as  well as  some simulators  developed to
optimise and test these proposals.
\end{abstract}

\section{Introduction}

The PDHS of  Gaia acquires the data coming from  the CCD focal planes,
selects and prioritizes them, encodes and compresses them  and finally
generates  the  corresponding  source  packets  to  be  fed  into  the
telemetry  stream.   We  have  developed  a proposal  for  its  global
operation, including the main modules  and the data flux between them.
For  the  moment we  have  focused on  the  Astro
instrument,  for which we  describe a  possible implementation  of its
video processing units. This proposal, however, can be extended to the
Spectro instrument.

Our  work  includes   not  only  the  system  design,   but  also  the
specification of the many operations  to be performed on board.  These
operations include an optimal codification  of timing data, as well as
an  optimal transmission  scheme fulfilling  the ESA  packet telemetry
standard. The main guidelines  for an  optimal data compression system 
are also described.  These guidelines will be the key  for fitting the
huge amount of  science data into the limited  downlink. A global view
of the  overall data  path is finally  discussed, from the  on board
instruments to the on ground storage.

It  is worth  noting  at this  point  that besides  these designs  and
specifications  we have  also developed a set of  simulation tools for
determining the optimal codification  parameters and for verifying the
reliability  of the telemetry  and data  compression systems.   One of
these  simulators  is  designed   as  a  large  software  application,
receiving  the output  of  some Gaia  simulators,  simulating all  the
telemetry and data compression  system, and returning the science data
to be fed into the data base an processing system.

\section{Design Proposals}

\subsection{Payload Data Handling System}

All  the science data  flux within  the spacecraft  is managed  by the
PDHS, from  the instruments to the communications  system. The PDHS
must be optimized and designed  as a pipeline, capable of concurrently
processing the huge  amount of data at its  several stages. This turns
out to  be crucial  because on average  about 200 stellar  objects per
second will  be measured (and processed), thus  implying internal data
fluxes of some hundreds of Mbps.

\begin{figure}[h]
\begin{center}
\leavevmode
\centerline{\epsfig{file=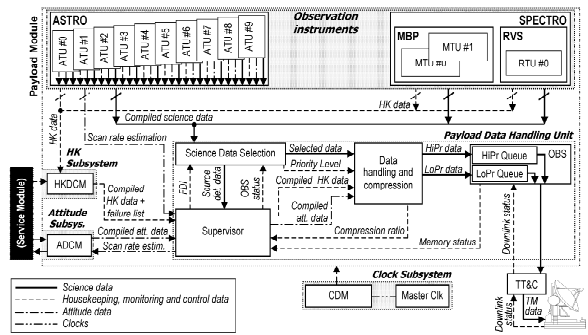,width=8.0cm}}
\end{center}
\caption{Overview of the PDHS of Gaia as proposed in our work.}
\label{fig:PDHS}
\end{figure}

\begin{figure}[t]
\begin{center}
\leavevmode
\centerline{\epsfig{file=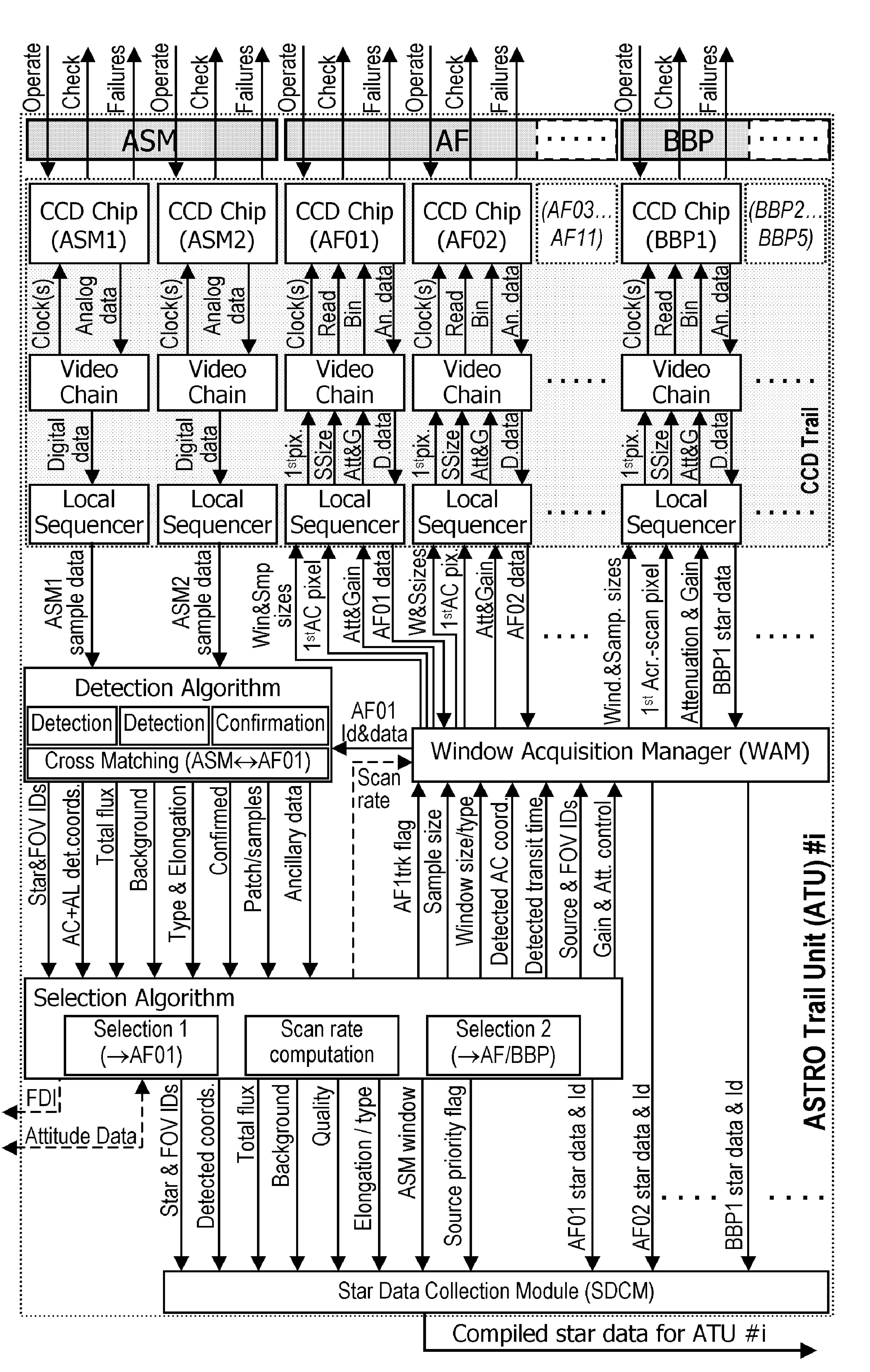,width=8.0cm}}
\end{center}
\caption{ Possible implementation of an Astro Trail Unit.}
\label{fig:ATU}
\end{figure}

The overall  design of  the PDHS is  shown in  Fig.~\ref{fig:PDHS}. We
propose  to deploy  Astro in  10 identical  sub-modules which  we name
Astro  Trail Units (ATUs)  operating in  parallel.  Fig.~\ref{fig:ATU}
shows a possible implementation of the Astro Trail Units. Each of them
will manage  the measurement of  stars transiting over  its associated
trail of CCDs.   Video chains and local sequencers  are used for this,
which not only combine the digital data from a single star measurement
but  also  operate  as  interfaces  between  high-level  commands  and
CCD-level  commands \citep{BCN004}.   The  window acquisition  manager
(WAM)  is in  charge  of  commanding this,  for  which an  acquisition
protocol has  been developed.  Unnecessary delays are  avoided and the
operation  is  pipelined.   The  detection  and  selection  algorithms
indicate to the  WAM which  sources  must be measured and  with  which
sampling scheme option.

The use of  a source priority flag is  also proposed \citep{BCN003} in
order to ensure that low-priority data may be easily discarded during
downlink shortages. Another recommended  flag, the Field Density Index
(FDI), would ease the control of field-dependant PDHS operations.  All
of these  data selection  procedures will be  executed by  the science
data selection  module. Afterwards, these pre-selected  raw data shall
be coded in  an optimised way in order  to avoid unnecessary telemetry
occupation. This will be the task of the Data Handling and Compression
module, which will also  include an optimised data compression system.
Finally, the  compressed and packetised data will  be stored on-board,
waiting  to be  transmitted during  the next  contact with  the ground
station.

\subsection{Optimised Time Data Codification}

\begin{figure}[t]
\begin{center}
\leavevmode
\centerline{\epsfig{file=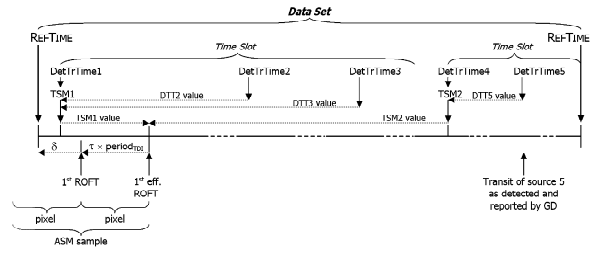,width=8.0cm}}
\end{center}
\caption{ Proposed timing scheme for Gaia.}
\label{fig:TimingScheme}
\end{figure}

Some of the critical data  generated by the instruments include timing
data,  which  can  produce  an  important  telemetry  occupation  and,
therefore,  their codification  must be  optimised. For  this, science
data  are grouped  in data  sets of  1 second  length  (in measurement
time), as already  assumed in the baseline. In  order to optimise even
more this  scheme, we propose to  partition every data  set in several
time slots,  in such  a way that  less bits  are required to  time tag
every   measurement   \citep{BCN006}.    Figure~\ref{fig:TimingScheme}
illustrates this scheme.

\subsection{Telemetry Generation}

\begin{figure}[h]
\begin{center}
\leavevmode
\centerline{\epsfig{file=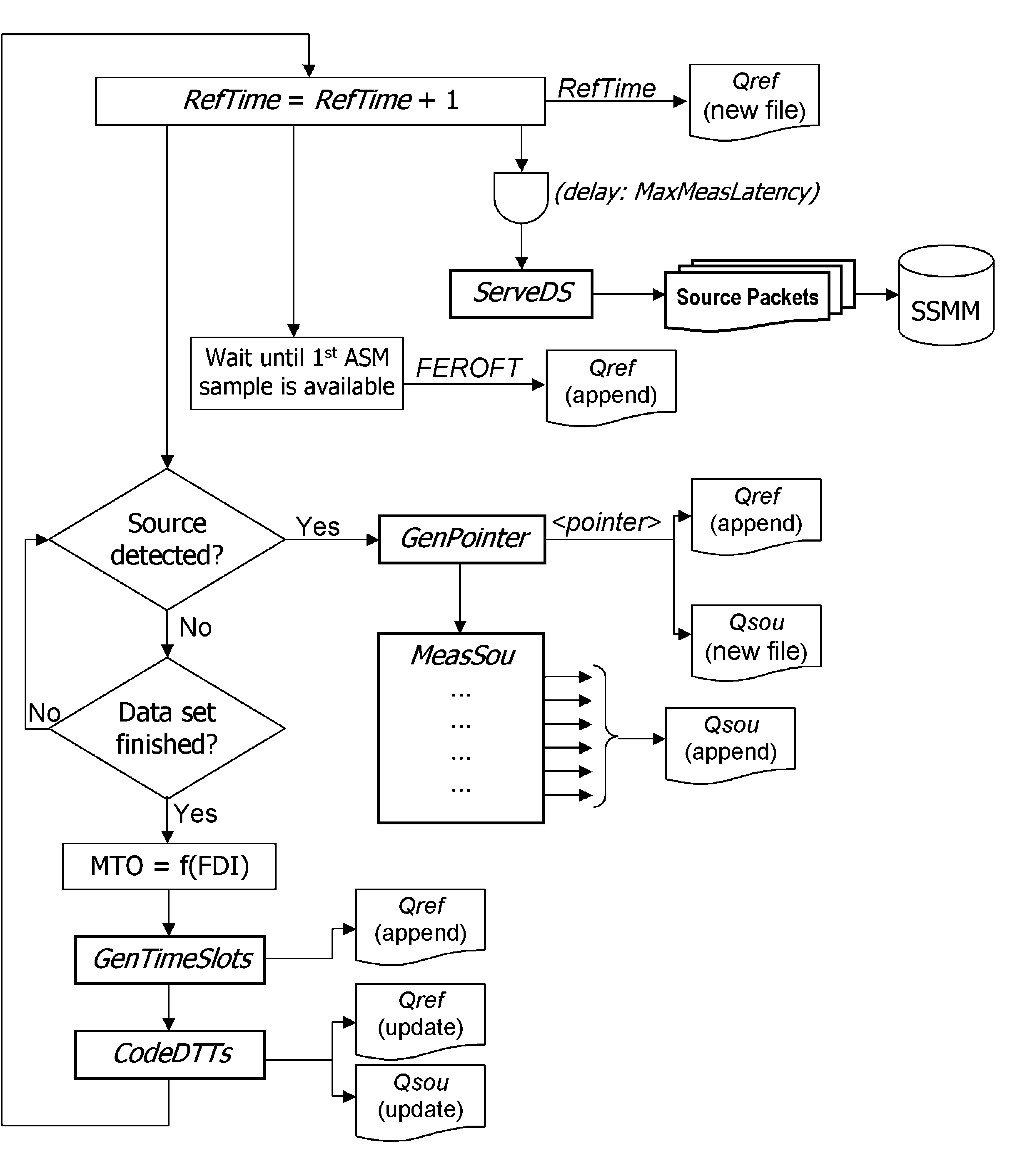,width=8.0cm}}
\end{center}
\caption{ Implementation guidelines for our optimized codification scheme.}
\label{fig:TMimplem}
\end{figure}

We  have devised a  codification scheme  \citep{BCN006} that  not only
implements our  optimised timing scheme, but also  fulfills the Packet
Telemetry  standard defined  by ESA.  Source Packets  are generated
accordingly  to  our  optimized codification  guidelines,  dynamically
adapting  to the  current  observation conditions.  The  core of  this
adaptive  system is  the {\sl Maximum  TSM Offset}  (MTO)  flag, which
indicates the length  of every time slot in which  we partition a data
set. Also, security systems have been introduced in order to avoid any
decoding    error.     Figure~\ref{fig:TMimplem}    illustrates    our
implementation proposal  for this optimized  and adaptive codification
system.

\subsection{Communications}

\begin{figure}[t]
\begin{center}
\leavevmode
\centerline{\epsfig{file=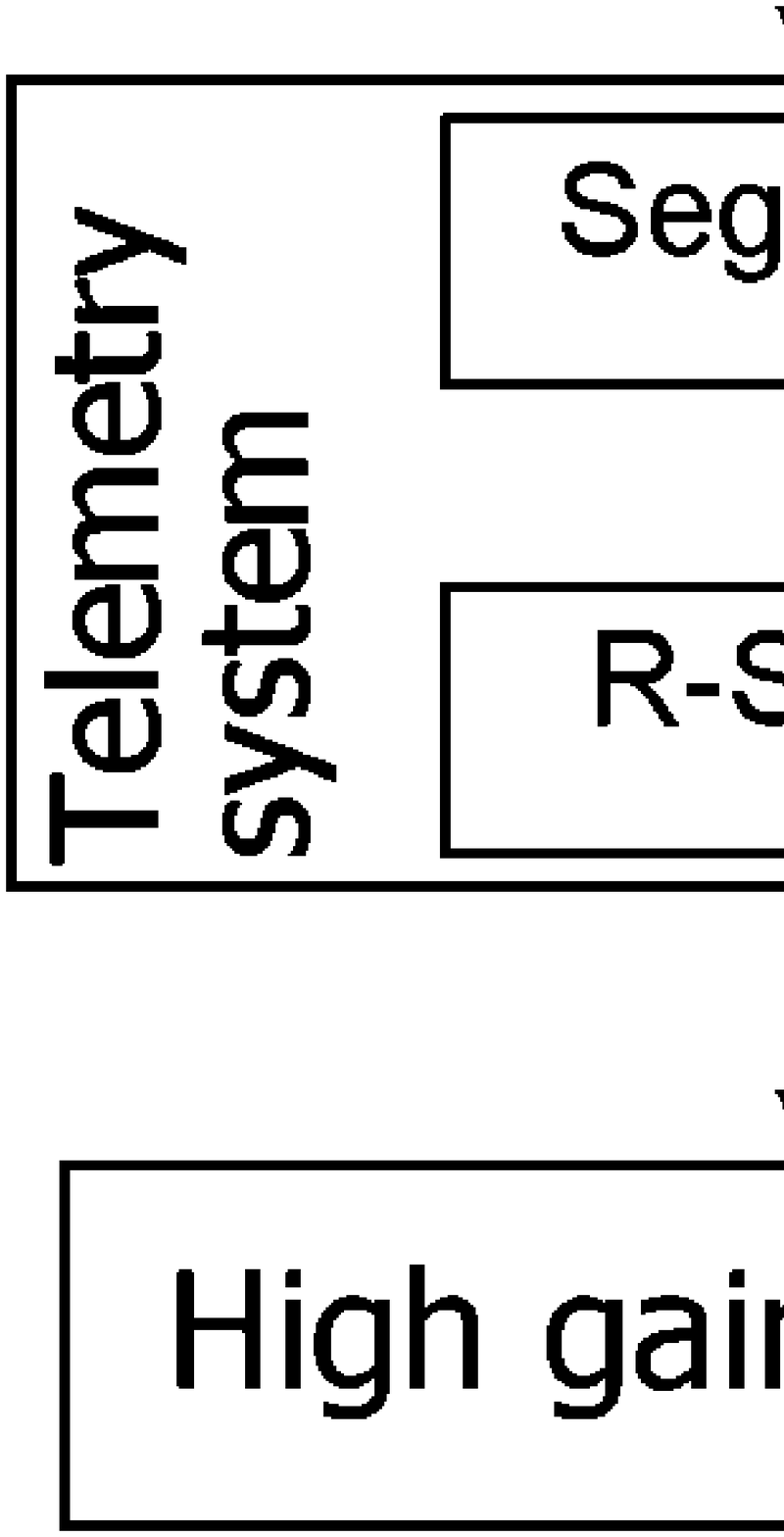,width=8.0cm}}
\end{center}
\caption{ Communication layers including our data handling proposals.}
\label{fig:CommLay}
\end{figure}

As ilustrated in  Figure~\ref{fig:CommLay}, our approach executes some
operations in an order different  of the usual one, packeting the data
{\sl before}  they are compressed.  By using this procedure  we ensure
that we keep each block of sources identified, making possible the use
of optimized  source packets and  making easier the  data priorisation
while dumping the on board storage to the ground station.

We   have  also  proposed   an  improved   channel  coding   for  Gaia
\citep{BCN008},  in  order to  decrease  the  minimum elevation  angle
required  to establish the  communication. This  minimum angle  in the
baseline is 10$^{\circ}$ above the horizon, while we proposed to reach
as low  as 5$^{\circ}$.  This could be achieved by adding  more correcting
codes within  the source packet  structures, which would  decrease the
codification efficiency  but only during  the 5$^{\circ}$-10$^{\circ}$
interval.  An adaptive system has been simulated with excellent results.

\section{Simulators}

\subsection{Optimisation of the Time Data Codification}

\begin{figure}[h]
\begin{center}
\leavevmode
\centerline{\epsfig{file=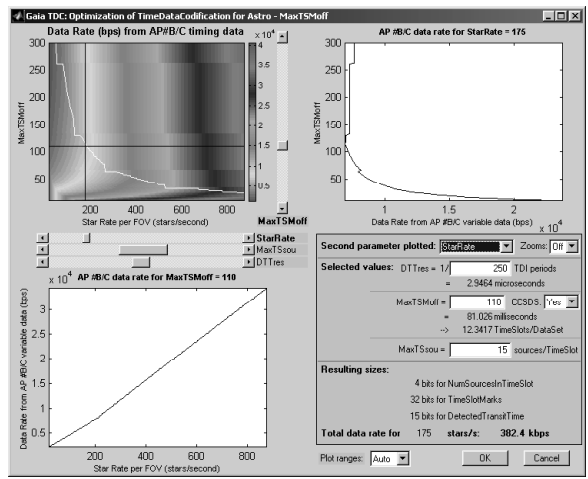,width=8.0cm}}
\end{center}
\caption{ Optimisation software for the codification parameters.}
\label{fig:StatTDCopt}
\end{figure}

\begin{figure}[h]
\begin{center}
\leavevmode
\centerline{\epsfig{file=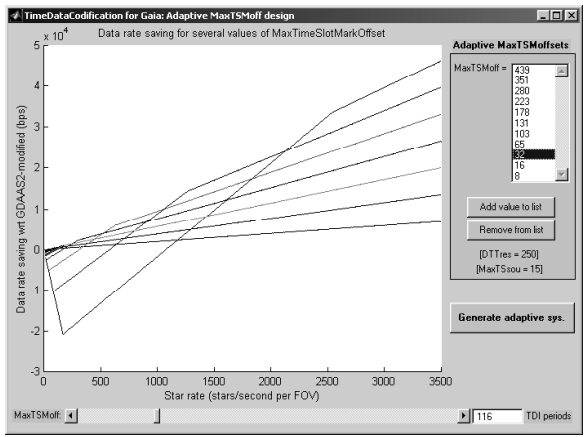,width=8.0cm}}
\end{center}
\caption{  Dynamic adaptation  of the  codification parameters  to the
         observation conditions.}
\label{fig:AdaptTDCopt}
\end{figure}

The  reliability  of our  optimised  proposal  for  the timing  scheme
depends  on  several  parameters.  We  have developed  a  software  to
simulate  the average  telemetry  data  rate as  a  function of  these
parameters     \citep{BCN006}.       The     snapshot     shown     in
Figure~\ref{fig:StatTDCopt}   shows    a   static   determination   of
codification parameters,  while Figure~\ref{fig:AdaptTDCopt} shows the
adaptive coding optimiser.  It led us to a dynamic  coding of the time
depending on the observed field density. The telemetry saving achieved
(only with this codification  system, that is, without any compression
yet) is  about 1.2  Kbps in average,  reaching up  to some 10  Kbps in
crowded fields. Our  simulator also offers an estimate  of the average
data rate, which is about 1.2 Mbps including both Astro fields of view
(without compression).

\subsection{Telemetry CODEC}

In   order  to   obtain   more  accurate   telemetry  simulations,   a
{\it~Telemetry  CODEC}  (coder/decoder)  software is  being  developed
\citep{BCN011}.   A preliminary  implementation of  this  software has
already been  successfully tested, receiving  realistic data generated
by GASS  (the Gaia System Simulator)  and converting it  to raw binary
data.  In this way, we can obtain realistic telemetry curves, based on
realistic star counts.

This  software, however,  is conceived  as a  large,  dynamic software
application, capable of receiving  data from different data simulators
and  performing complex operations  on them.  Furthermore, it  will be
configured with XML  files, which not only avoids  the modification of
the code for adapting to different telemetry models but will also make
possible to fulfill forthcoming XML telemetry standards. This software
is  currently being  coded, but  preliminary tests  are  also offering
satisfactory  results. Even  the  data compression  software shall  be
integrated in the Telemetry CODEC, as well as statistical studies that
will determine the telemetry occupation and compression ratio achieved
on every data field type.

\subsection{Data Compression}

One of the  callenges in Gaia is to transmit the  large amount of data
from  the  satellite to  the  ground  station.  Preliminary  estimates
\citep{UweTM} show  an average data  generation rate of about  5 Mbps,
while sustained downlink  capability will be about 1.5  Mbps. This, in
turn, implies that the data must  be compressed a factor of 3 or more,
using lossless algorithms whenever possible.  This is, in fact, a true
challenge, since tests with  standard compression systems do not offer
more than  1.5 in the  best of the  cases.  Therefore, a  tailored and
optimised system must be developed for Gaia.  We have devised a set of
compression techniques, most  of them based on PSF  and Galaxy models,
as  well as  differential coding,  predictors and  stream partitioning
\citep{PortellPFC,BCN001,BCN011}.      These    will     operate    as
pre-compressors, after which the application of standard systems offer
much  better  results,   as  shown  in  Table~\ref{tab:compr}.   These
preliminary simulations, obtained with  GASS v2.1 data, reveal that we
are in the good direction since a factor of 2.4 is completely feasible
by using our methods. Further detailed studies and developments should
bring us to the desired target of 3 (or even more) in a near future.

\section{Conclusions}

We have proposed a set of designs for the payload data handling system
of Gaia, including an overview of the system, its modules and the main
data flux between them. Many  of these modules have also been defined,
whether as another  set of submodules (as the Astro  trail unit) or as
operation guidelines  (such as the data handling  modules). The latter
include accurate proposals for the timing and transmission schemes, as
well as for an optimised data compression. All of these proposals take
into account the  latest design of Gaia and the  need for an optimised
system,  in  terms  of  hardware requirements,  processing  speed  and
reduced telemetry occupation.

\begin{table}[t]
  \caption{Data compression ratios obtained with different methods.}
  \label{tab:compr}
  \begin{center}
    \leavevmode
        \begin{tabular}[h]{lcc}
      \hline \\[-5pt]
      Pre-compressor & Compressor & Best ratio \\[+5pt]
      \hline \\[-5pt]
      None & gzip  & 1.35 \\
      None & bzip2  & 1.49 \\
      Adaptive differential & None  & 1.60 \\
      Adaptive differential & gzip & 1.78 \\
      Stream partitioning & gzip & 2.35 \\
      Stream partitioning & bzip2 & 2.48 \\
      \hline \\
      \end{tabular}
  \end{center}
\end{table}

Many  of  these proposals  depended  on  parameters  which had  to  be
determined for  a realistic case. For  this, we have  also developed a
set  of simulation  tools  which include  the  optimisation of  timing
parameters,  the  generation   of  realistic  telemetry  streams  from
simulated  science  data,  and  the  compression  of  these  telemetry
data.  Although some of  these software  applications are  still being
developed, their  preliminary results  are very encouraging.  Our data
compression simulator  is specially interesting, since  it is offering
the highest ratio currently achieved on realistic Astro data.

\section*{Acknowledgements}

This   work  has   been  partially   supported  by   the   MCYT  grant
AYA2002--4094--C03--01, by  the European Union FEDER funds  and by the
CIRIT.


\begin{thebibliography}{}
 
\bibitem
 [\protect\astroncite   {Geijo  et~al.}{2004}]{BCN008}   Geijo,  E.M.,
 Portell,  J., Garc\'{\i}a--Berro,  E., Luri,  X., Lammers,  U., 2004,
 Improved  channel coding  for longer  contact times  with  Gaia, Gaia
 internal technical report (GAIA-BCN-008)
 
 \bibitem
 [\protect\astroncite {Lammers}{2004}]{UweTM}  Lammers, U., 2004, Gaia
 telemetry  rate simulations: a  first look  at the  complete picture,
 Gaia internal technical report (GAIA-UL-008)
 
 \bibitem
 [\protect\astroncite {Portell}{2000}]{PortellPFC}  Portell, J., 2000,
 Dise\~no  del  enlace  de  datos  para SIXE  (Spanish  Italian  X-ray
 Experiment), in Spanish, Master Thesis, UPC
 
 \bibitem
 [\protect\astroncite  {Portell  et~al.}{2001}]{BCN001}  Portell,  J.,
 Garc\'{\i}a--Berro,  E.,  Luri,  X.,  2001, Flux  data  codification:
 proposals  of simple  codification schemes,  Gaia  internal technical
 report (GAIA-BCN-001)
 
 \bibitem
 [\protect\astroncite {Portell}{2001}]{BCN003} Portell, J., 2001, Some
 ideas  for   GIBIS  and   beyond,  Gaia  internal   technical  report
 (GAIA-BCN-003)
 
 \bibitem
 [\protect\astroncite  {Portell  et~al.}{2003}]{BCN004}  Portell,  J.,
 Garc\'ia-Berro,  E., Luri,  X.,  2003, Proposals  of  a payload  data
 handling unit and internal  data flux, Gaia internal technical report
 (GAIA-BCN-004)
 
 \bibitem
 [\protect\astroncite  {Portell  et~al.}{2004a}]{BCN006} Portell,  J.,
 Garc\'{\i}a--Berro,  E.,  Luri,  X.,  2004, Timing  and  transmission
 schemes for Gaia, Gaia internal technical report (GAIA-BCN-006)
 
 \bibitem
 [\protect\astroncite  {Portell  et~al.}{2004b}]{BCN011} Portell,  J.,
 Luri,  X., Garc\'{\i}a--Berro,  E., 2004,  Definition of  a Telemetry
 CODEC, Gaia internal technical report (GAIA-BCN-011)
 
 
 \end{thebibliography}
\end{document}